\documentclass[useAMS,usenatbib,usegraphicx]{mn2e}
\usepackage{fixltx2e} 
\usepackage{url}    
\usepackage[svgnames]{xcolor}  %
\newcommand{\arcdeg}{\mbox{$^\circ$}}
\newcommand{\uv}{\mbox{$u$-$v$}}

\newcommand{\pyear}{\mbox{\% yr$^{-1}$}}

\newcommand{\Jb}{\mbox{Jy beam$^{-1}$}}
\newcommand{\uJb}{\mbox{$\mu$Jy beam$^{-1}$}}
\newcommand{\invassq}{\mbox{arcsec$^{-2}$}}
\newcommand{\Jyas}{\mbox{Jy arcsec$^{-2}$}}
\newcommand{\PSRB}{\mbox{PSR B0531+21}}

\newcommand{\AD}{\mbox{\textsc{ad}}}

\newcommand{\phcms}{\mbox{photons cm$^{-2}$ s$^{-1}$}}
\newcommand{\lesssim}{\mbox{\raisebox{-0.3em}{$\stackrel{\textstyle <}{\sim}$}}}
\newcommand{\gtrsim}{\mbox{\raisebox{-0.3em}{$\stackrel{\textstyle >}{\sim}$}}}

\title[Crab: Radio variability and gamma-ray flares]{The variability
  of the Crab Nebula in radio: No radio counterpart to gamma-ray
  flares}
\author[Bietenholz et al]{M. F. Bietenholz$^{1,2}$, Y. Yuan$^3$,
  R. Buehler$^{4}$, A. P. Lobanov$^{5,6}$, and R. Blandford$^3$\\
$^1$Hartebeesthoek Radio Observatory, PO Box 443, Krugersdorp,
1740, South Africa \\
$^2$Department of Physics and Astronomy, York University, Toronto,
M3J~1P3, Ontario, Canada \\
$^3$Kavli Institute for Particle Astrophysics and Cosmology, Stanford 
University, Menlo Park, CA 94025, US\\ 
$^4$Deutsches Elektronen Synchrotron (DESY), Platanenalee 6, 15738, Zeuthern, Germany \\
$^5$Max-Planck-Institut f\"ur Radioastronomie, Auf dem H\"ugel 69,
53121 Bonn, Germany\\
$^6$Institut f\"ur Experimentalphysik, Universtit\"at Hamburg, Luruper Chaussee 149, 22761, Hamburg, Germany \\
}
\begin{document}
\date{\today;  {\em Accepted to MNRAS}}
 
\maketitle
\label{firstpage}

\begin{abstract}
We present new Jansky Very Large Array (VLA) radio images of the Crab
Nebula at 5.5 GHz, taken at two epochs separated by 6 days about two
months after a gamma-ray flare in 2012 July.  We find no significant
change in the Crab's radio emission localized to a region of $<2$
light-months in radius, either over the 6-day interval between our
present observations or between the present observations and ones from
2001\@.  Any radio counterpart to the flare has a radio luminosity of
$\lesssim 2 \times 10^{-4}$ times that of the nebula. Comparing our
images to one from 2001, we do however find changes in radio
brightness, up to 10\% in amplitude, which occur on decade timescales
throughout the nebula.  The morphology of the changes is complex
suggesting both filamentary and knotty structures. The variability is
stronger, and the timescales likely somewhat shorter, nearer the
centre of the nebula.  We further find that even with the excellent
\uv~coverage and signal-to-noise of the VLA, deconvolution errors are
much larger than the noise, being up to 1.2\% of peak brightness of
the nebula in this particular case.
\end{abstract}


\section{Introduction}
\label{sintro}

The Crab Nebula is one of the most intensely studied objects in
astrophysics, yet it retains the power to surprise us.  It is the
remnant of a supernova explosion in the year \AD\ 1054, which was
witnessed by Chinese and other astronomers.  The presently visible
remnant is bright at all observable wavelengths.  It has, in
particular, long been observed in the radio, where it is one of the
brightest sources at GHz frequencies \citep[see, e.g.,][and references
  therein]{Hogg+1969, Wilson1972c}.
It also contains one of the first known pulsars, \PSRB.  The Crab
Nebula is the prototype of a pulsar-powered nebula, commonly known as
a pulsar wind nebula (PWN), where the rotational energy lost by the
pulsar as it spins down powers presently visible nebula
\citep[see][for recent reviews]{BuehlerB2014, Hester2008}.  The energy
input from the pulsar, which emerges in the form of a wind of magnetic
field and relativistic particles has inflated a bubble of relativistic
fluid.  Since the pulsar spindown is quite regular, the rate of this
energy input is quite steady, although it decreases slowly, 
in the case of the Crab at  $\sim$0.14\pyear.

A reason to re-observe the Crab in the radio as well as at other
wavelengths occurred when it was recently discovered that the Crab
produces substantial flares in gamma-ray emission. Several such flares
have now been observed, where the emission at photon energies
$>100$~MeV increases by more than a factor of two on timescales of
days \citep[e.g.][]{Buehler+2012,Striani+2011,Tavani+2011}.
These flares are not yet well understood --- see discussions in e.g.\
\citet{Arons2012, Bykov+2012, LyutikovBM2012, Lyutikov2014,
SturrockA2012}, and \citet{KomissarovL2011}, although they likely
involve regions moving relativistically towards us and/or perhaps
magnetic reconnection.  While there are occasionally small sudden
changes in the spindown rate of the pulsar known as glitches,
\citet{Espinoza+2010} and \citet{Mickaliger+2012} have shown glitches
do not seem to occur in conjunction with the gamma ray flares.

On 2012 July 3, the Fermi Large Area Telescope (Fermi-LAT) detected a
new flare from the Crab \citep{Ojha+2012} in gamma-rays, with the flux
at energies $>100$~MeV having increased to $(5.5 \pm 0.7) \times
10^{-6}$ \phcms\ from levels near its long-term average level of $2.75
\times 10^{-6}$~\phcms\ \citep{Nolan+2012}.  The daily average flux
reached a peak of $(6.2 \pm 0.8) \times 10^{-6}$~\phcms, over twice
the long-term average value, but by July 8 it had decayed back to the
long-term average level (preliminary values from our reduction,
Buehler).  Note that the above flux values are the totals, consisting
of the combined flux from both the pulsar and the nebula.  The average
flux from the nebula is only $0.6 \times 10^{-6}$~\phcms. 

The short timescales of the flares imply that the emitting region is
quite small: unless there is ultrarelativistic beaming, the flare
emission must come from a region of $\lesssim 3 \times 10^{-4}$~pc,
corresponding to $\lesssim 0\farcs03$ in size \citep[and references
  therein]{Abdo+2011, BuehlerB2014}.
A similar conclusion was reached by modelling the combined gamma-ray,
X-ray and radio spectrum \citep{LobanovHM2011, MeyerHZ2010}.

Since the flaring component of the emission is not pulsed, it is
thought to originate in the nebula rather than the pulsar
\citep{Abdo+2011, Buehler+2012}.  Together with the short timescales,
this implies that it must be synchrotron emission, arising from
electrons with energies of order $10^{15}$~eV. 
In gamma rays, the nebula therefore brightened by almost a factor of 6
during the July 2012 flare.
Although this flare was slightly less luminous than the earlier
flares, it still represents a very energetic event, as well as only
the fourth time such high fluxes had been observed in four years of
Fermi observations.

The resolution of Fermi-LAT is on the order of 1\arcdeg, therefore the
gamma-ray emission cannot be localized within the nebula (which is
about $7\arcmin \times 5\arcmin$ in extent).
Telescopes from infrared to X-ray observed the Crab during or shortly
after previous flares and found no unusual emission \citep[and
references therein]{Abdo+2011, Buehler+2012}. 

Several locations in the inner nebula have in fact been suggested as
the location of the gamma-ray flares.  They include the ``inner knot''
first identified by \citet{Hester+1995}\footnote{This feature was
called ``Knot 1'' in \citet{Hester+1995}}, as well as the ``anvil''
region suggested by \citet{Tavani+2011}.  As of yet, however, the
location of the gamma-ray flaring within the nebula has not yet been
identified. 

Since radio emission is often associated with gamma-ray emission, and
since arcsecond resolution is easily achievable in the radio, radio
observations seem a natural choice to attempt to pinpoint the location
of the flare activity.  In more energetic flares, the spectrum of the
flaring part of the gamma-ray emission appears to have a cutoff below
photon energies of $\sim$200~MeV, but the nature of the spectrum of
the less energetic flares, like the one of 2012 July, is not clear
\citep{BuehlerB2014}.
Although any extrapolation of the spectral energy distribution seen in
the gamma-ray flares to radio frequencies (photon energies of order
$10^{-5}$~eV), seems dangerous, the source of the energy seems likely
to be the magnetic field, and one might expect disturbances in the
magnetic field to produce changes in the radio-frequency synchrotron
emission.  Furthermore, the electrons responsible for the flaring
gamma-ray emission should certainly still be energetic enough to
produce radio emission long after the flare event.  

The very energetic electrons producing the gamma-ray emission have
very short lifetimes, as they rapidly lose their energy due to
synchrotron losses.  By contrast, the lifetimes of the less energetic
electrons producing radio emission is generally longer than the age of
the nebula.  Therefore, even though the high-energy emission is
short-lived, any response to it in particles emitting at radio
frequencies would be adiabatic, and any corresponding radio emission
should be discernible until the particles diffuse into the body of the
nebula.  

In fact, the origin of the relativistic electrons producing the radio
synchrotron emission, i.e.\ those having energies $\lesssim 10$~GeV,
is a long standing puzzle.  These electrons have synchrotron lifetimes
long compared to the age of the nebula.  If the radio-emitting
electrons are continuously injected into the nebula along with the
higher energy ones, the total number of particles is larger than
produced by current theories \citep{Arons2012,HibschmanA2001}.
However, if they are not injected by the pulsar, but accelerated by
some other process in the nebula, then the existence of radio
features, called ``wisps'', associated with the termination shock
\citep{Crab-2004, Crab-2001, Crabwisp-1992} is a puzzle, as is the
continuity of the spectral energy distribution from the radio through
to the optical and above \citep{AharonianA1998}. The first of these
issues may not be a problem: \citet{Olmi+2014} have shown that even if
the radio-emitting electrons are produced in the body of the nebula,
their long lifetimes would allow them to diffuse essentially
everywhere in the nebula, and they would still be expected to be
present near the termination shock, and that variations of the
magnetic field near the shock would then naturally produce the radio
wisps. \citet{Komissarov2013} has shown that the radio-emitting
electrons could plausibly be accelerated by magnetic dissipation in
the vicinity of the filaments.

In any case, since the acceleration of the higher energy particles
responsible for the gamma-ray flare is likely to happen near the
pulsar, the detection of (or upper limit to) any radio counterpart to
the flares would shed light on the low-energy tail of the acceleration
mechanism responsible for the gamma-ray flares, and might therefore
also shed light on the more general process for the production of the
radio-emitting electrons.

Radio observations of the Crab in response to gamma-ray flares have
been previously undertaken.  Firstly, \citet{LobanovHM2011} obtained
1.4-GHz VLBI observations, most sensitive to features $<0\farcs1$ in
size, 44~d after the flare of 2011 Sep.  They detected two weak
features, called knots C1 and C2, which they tentatively identified as
possible radio counterparts to the gamma-ray flares.  The flux density
of these knots C1 and C2 was 0.5 and 0.4 mJy and their distance from
the pulsar 5\farcs4 and 1\farcs6, respectively.
Secondly, \citet{Weisskopf+2013} obtained a series of
Karl G. Jansky Very Large Array (VLA), as well as {\em Chandra} X-ray
and Keck optical, observations at $t = 3$ to 89~d following the strong
2011 April flare.  Their radio observations had arcsecond resolution.
They report no compact radio emission features $> 0.5$~mJy except for
the pulsar, although their radio observations were not sensitive
enough to detect C1 or C2 at the flux densities found by
\citet{LobanovHM2011}.

In response to the 2012 July gamma-ray flare, we therefore undertook
Target of Opportunity observations of the Crab with the VLA to search
for radio emission in response to the gamma-ray flare, and confirm the
tentative detection of \citet{LobanovHM2011}.  The 2012 July flare
fortuitously occurred while the VLA was in the B-array configuration
which is well suited to high-resolution imaging of the Crab, giving
arcsecond resolution at 5~GHz.

A confirmed detection of radio emission in response to a gamma-ray
flare would localize the flaring region in the complex structure of
the nebula.  A detection, or even a non-detection with a sensitive
upper limit, would provide an important constraint on the broad-band
spectral energy distribution of the emitting material.  

Many of the current models of the flares \citep[e.g.][]{Lyutikov2014,
  Cerutti+2013, Clausen-BrownL2012}
involve reconnection events, which would produce quasi-monoenergetic
particle distributions, which would give rise to synchrotron emission
with a spectrum $\propto \nu^{1/3}$ below the peak frequency and which
would be unobservably faint in the radio.  However, the large energy
release would almost certainly cause disturbances in the magnetic
field, which should be made visible in radio via the synchrotron
emission of long-lived radio-emitting electrons not specifically
accelerated by the flaring event.  

To our knowledge, there have been no calculations of the expected
radio brightness changes in response to a gamma-ray flare.  We perform
such a calculate below (Section \ref{sdiscuss}) and determine that the
radio brightness changes may well be too small to be detectable in our
observations.  However, this calculation is very model-dependent, and
was only performed after we had obtained the target-of-opportunity
radio observations which are the subject of the present paper.
Particularly given the tentative detection of \citet{LobanovHM2011} we
consider that our new observations were well motivated.

\section{Observations and data reduction}
\label{sobs}

We obtained two sessions of VLA observations (observing code 12A-486)
of the Crab
with midpoints 2012 August 20.5 and 26.5 UT, spaced 6.0 days apart, or
49 and 55 days after the onset of a gamma-ray flare. The observations
employed a bandwidth 2048 MHz around a central frequency of 5503 MHz,
spanned a total of 5 hours per session, and, as already mentioned,
were carried out with the array in the B configuration, resulting in
baseline lengths between $\sim$2 and $\sim$250~K$\lambda$.  The flux
density scale was calibrated using observations of 3C~48 and 3C~147,
using the Perley-Butler (2010) coefficients, and we used QSO
J0559+2353 (QSO B0556+238) as a phase-reference source.

Observations in the B array configuration at 5 GHz do not provide
sufficient information at low spatial frequencies to reconstruct the
entire nebula.  In particular, the observations contain no direct
information on spatial scales larger than $\sim$50\arcsec.  We employ
the strategy for obtaining reliable images of the Crab using only a
single array configuration devised by \citet{Crab-2001, Crab-2004},
and we repeat a brief description here.  The strategy involves
employing maximum-entropy deconvolution using a support made from the
B, C and D array configuration observations taken in 1987 to 1988
\citep{Crab-Faraday,Crab-jet}.  The support was scaled spatially to
account for the expansion of the synchrotron nebula of 0.13\pyear\
\citep{Crab-expand}, and in brightness to account for the long-term
decay of the nebula\footnote{We take a weighted average of the
decay-rate values from \citet{Weiland+2011} at 22 and 30~GHz,
\citet{AllerR1985b} at 5~GHz and \citet{Vinyaikin2007} at 927~MHz; our
results are not sensitive to small changes in this decay rate.} at
0.202\%~year$^{-1}$.
The use of the same default for both sessions serves to make any
differences between our images at different epochs be only those that
are demanded by the data.

\section{Results}
\label{sresults}
\subsection{Radio images of the Crab}
\label{simages}

\begin{figure*}
\centering
\includegraphics[width=0.9\textwidth]{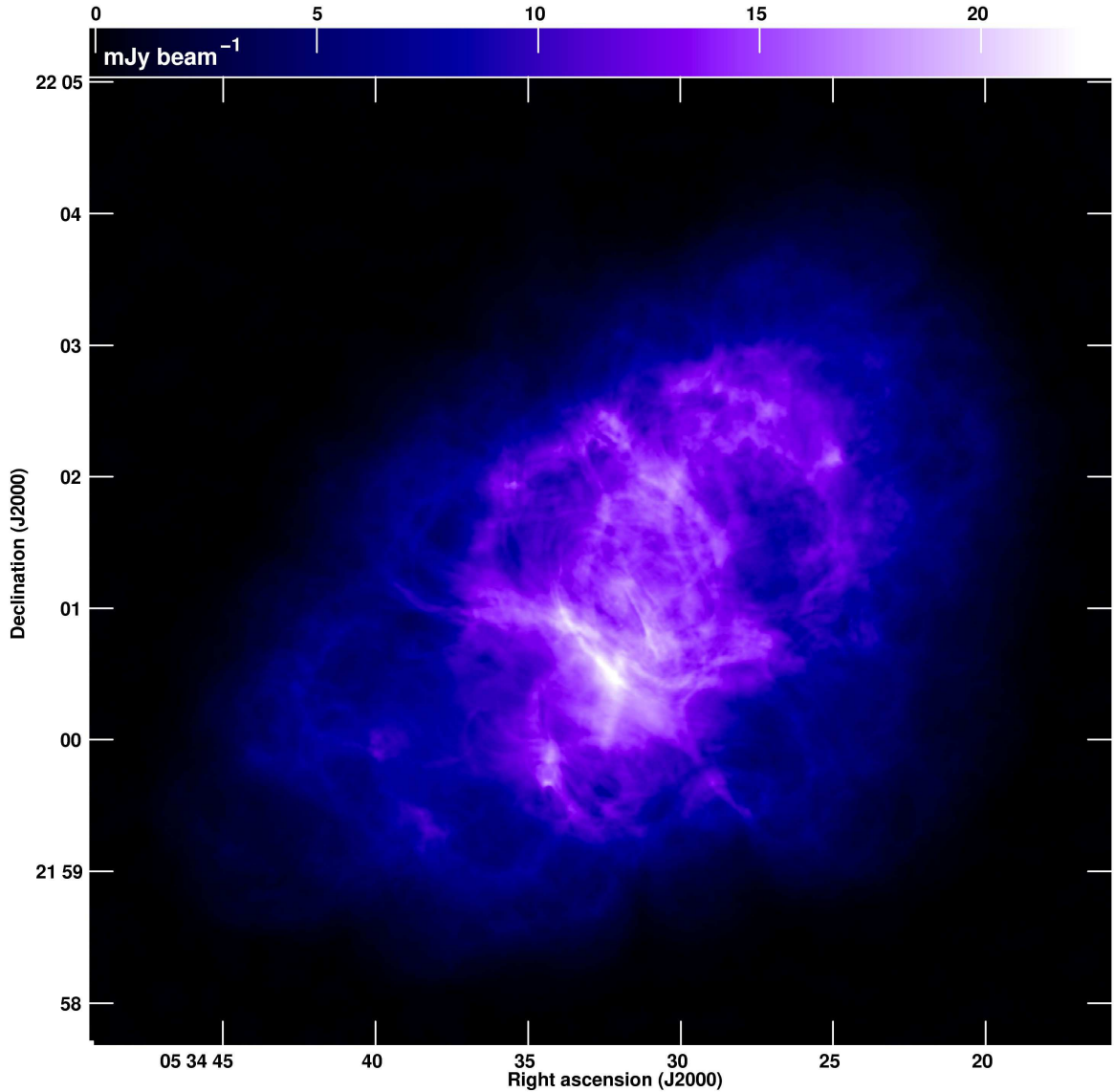}
\caption{An image of the Crab Nebula at 5.5 GHz on 2012 August 26\@.
  The FWHM of the convolving beam was $1.03\arcsec \times 0.99\arcsec$
  at p.a.\ $-66$\degr.  The
  peak brightness was 22.9~m\Jb, and the background rms was 28~\uJb.
  The total flux density recovered in the deconvolution was 592~Jy.
  Maximum entropy deconvolution was used, with an appropriately scaled
  default image made from the 1987 and 1988 multi-configuration VLA
  observations (see text for details).  Most of the structure on large
  spatial scales (\protect\gtrsim 37\arcsec) is derived from the
  default image
  rather than from the present observations. The image has been
  corrected for the effect of the VLA primary beam response.}
\label{fcrabimage}
\end{figure*}

\begin{figure}
\centering
\includegraphics[width=0.48\textwidth]{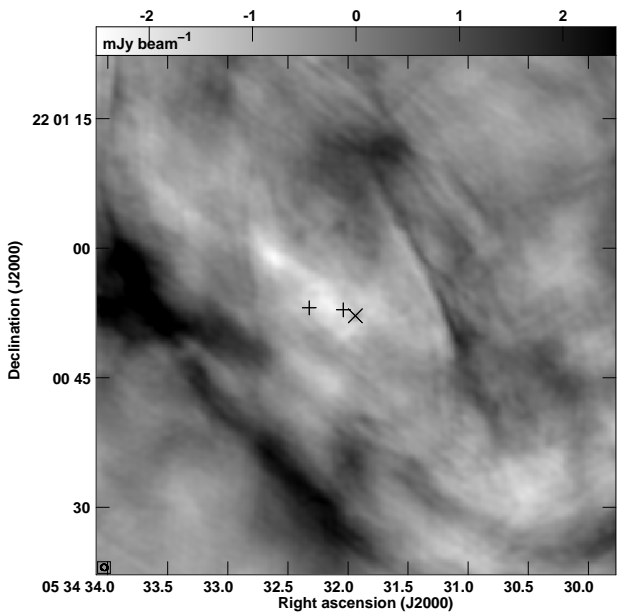}
\caption{A higher resolution 5.5-GHz image of the central region of
  the Crab on 2012 Aug.~26.  The image is made using the same
  visibility data as Fig.~\ref{fcrabimage}, but using uniform
  weighting for higher resolution, and multi-resolution CLEAN
  deconvolution.  Since no support information is used, this image is
  high-pass filtered, and consequently has a mean brightness near 0,
  but should accurately show the details at small-spatial scales.  The
  FWHM of the restoring beam ($0.80\arcsec \times 0.72\arcsec$ at
  p.a.\ $68$\arcdeg) is indicated at lower left.  We mark the pulsar,
  \PSRB, with ``$\times$'' and the knots C1 and C2 from \citet[see
    text \S~\ref{sdiscuss}]{LobanovHM2011} with ``+''.  The pulsar can
  be faintly discerned and has a flux density of $1.2 \pm 0.3$~mJy.}
\label{fcentre}
\end{figure}

We show the radio image of the Crab nebula at 5~GHz on 2012 August 26
in Figure~\ref{fcrabimage}.  Since the upgraded VLA has much wider
bandwidth than it did during the earlier observations of the Crab,
this image has both better image fidelity due to better coverage in
the Fourier-transform or \uv~plane, as well as lower noise, than any
previous radio images of the Crab.  In Figure \ref{fcentre} we show a
detail of the region around \PSRB.

\begin{figure*}
\centering
\includegraphics[width=0.9\textwidth]{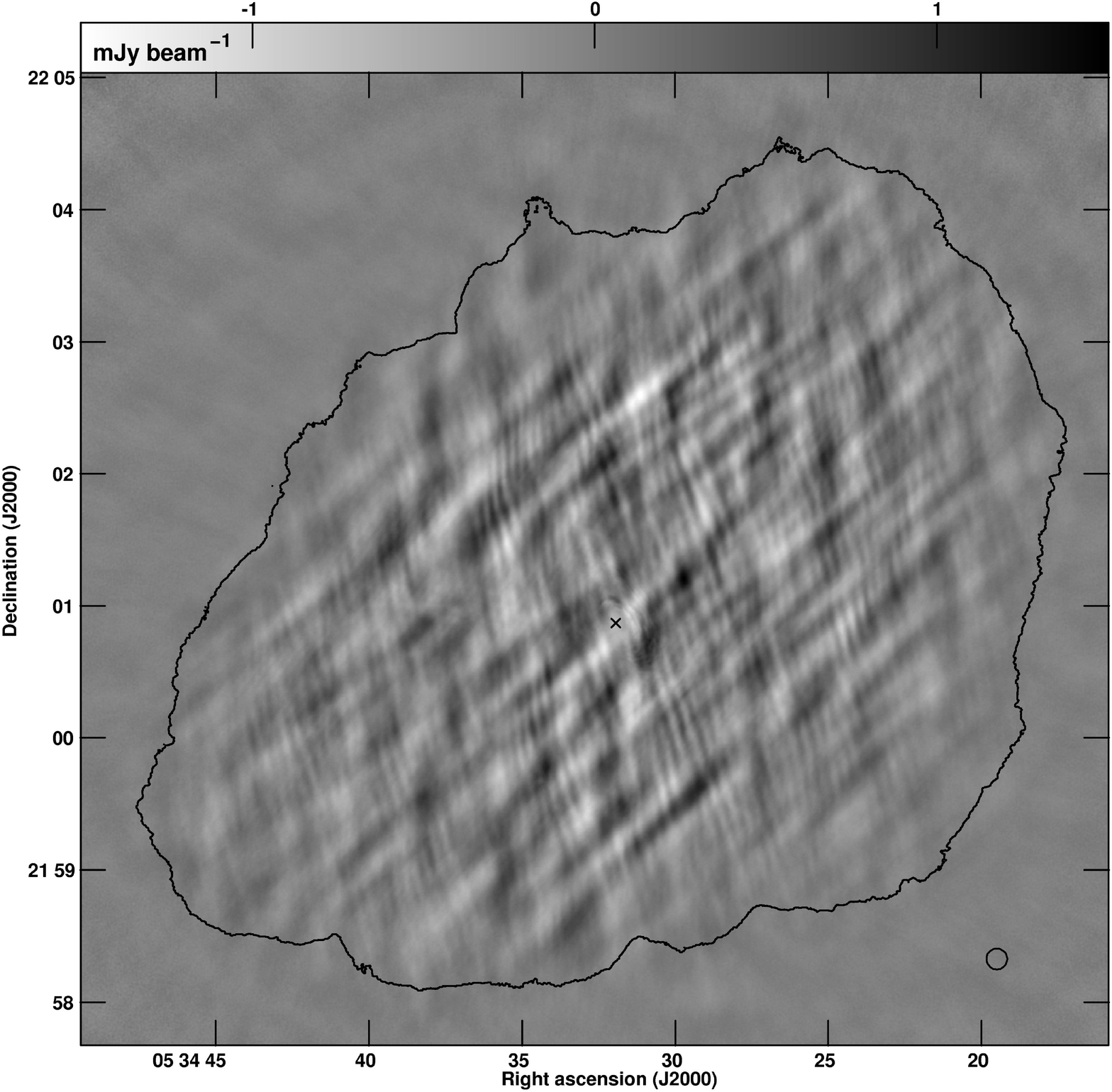}
\caption{A difference image of the Crab, showing the changes between
  2012 August 20 and 26\@ (with the former subtracted from the
  latter). The white
  ``$\times$'' in the centre gives the position of
  \PSRB\ \citep{LobanovHM2011}.  For reference, we also show the 1\%
  contour of the August 26 image.  The difference image has been
  high-pass filtered at 37\arcsec, since the observations were not
  sensitive to structure on larger scales.  Before forming the
  difference, both images were convolved to a common restoring beam
  size of $1.10\arcsec \times 1.00\arcsec$ at p.a.\ $-65$\arcdeg.  The
  off-source rms in the difference image in the region around the Crab
  was $\sim 41$~\uJb.  No isolated feature is visible which might
  correspond to the location of the gamma-ray flare event.  The circle
  at lower right shows a diameter of 9\farcs5, which is the diameter
  that a feature expanding isotropically with speed $c$ since the
  gamma-ray flare onset on 2012 July 3 would have at distance 2 kpc.}
\label{fdiff2012}
\end{figure*}

To determine whether there were any rapid changes in the radio images
due to the flaring activity, we formed a difference image between our
two epochs.  Since both epochs were deconvolved using the same default
image, they will be biased to be as similar as is allowed by the
visibility measurements, thus reducing the likelihood of spurious
difference features due for example to deconvolution errors.  Before
differencing, both images were convolved to a common restoring beam,
for which we chose a size of $1.10\arcsec \times 1.00\arcsec$ at
p.a.\ $-65$\arcdeg, marginally larger than the value fit to the dirty
beams of each individual epoch.

An interferometer is a spatial filter, so the VLA in the B-array at
our lowest frequency of 4.7~GHz can reproduce spatial scales smaller
than\footnote{VLA Observational Status Summary, Table 5;
  \url{https://science.nrao.edu/facilities/vla/docs/manuals/oss2014a}}
37\arcsec.  Accordingly, we high-pass filtered the difference image
with a Gaussian filter with 37\arcsec\ FWHM to remove any spurious
structure at lower spatial scales (note that our use of a common
support for the two epochs should have already served to minimize any
such differences).

\begin{figure}
\centering
\includegraphics[width=0.48\textwidth]{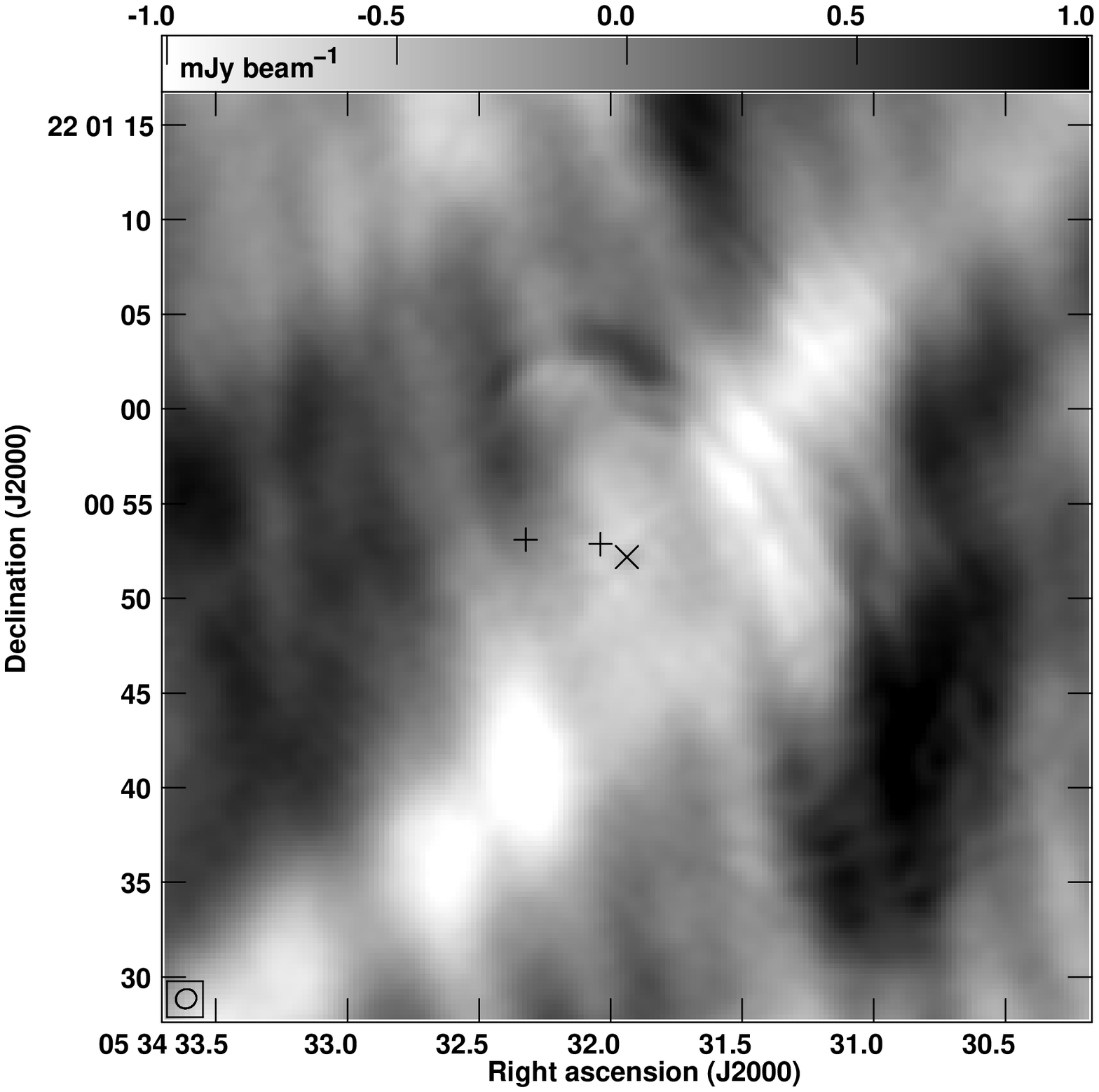}
\caption{A detail from Fig.~\ref{fdiff2012}, showing the region near
  the pulsar in the Aug.\ 26 $-$ Aug.\ 20 difference image.  The
  FWHM of the restoring beam
  ($1.10\arcsec \times 1.00\arcsec$ at p.a.\ $-65$\arcdeg) is
  indicated at lower left.  We mark the pulsar with ``$\times$'' and
  the knots C1 and C2 from \citet[see text
    \S~\ref{sdiscuss}]{LobanovHM2011} with ``+''.  The pulsar is
  faintly visible as a positive excursion, implying that the
  pulse-averaged flux density of the pulsar increased by
  $\sim$200~\uJb\ between the two dates. A series of ripples due to
  the outward motion of the elliptical wisps is also visible.}
\label{fdiffcentre}
\end{figure}

We show the resulting difference image in Figure~\ref{fdiff2012}, and 
a detail of the centre in Figure~\ref{fdiffcentre}.
The rms differences over the body of the Crab nebula were 0.26~m\Jb,
and the extrema of the difference image were $-1.48$ and $+1.56$~m\Jb,
with the area examined being $\sim$74,000 beam areas.  Expressed in
terms of the peak brightness of the individual images, which was
$\sim$25~m\Jb\ at this resolution, the peak-to-peak
difference was about $\pm 6$\% while the rms value over the body of
the nebula was 1.2\%.

The overall rippling in the difference image, as we will show in more
detail in \S~\ref{scaution} below, is mostly due to deconvolution
errors and likely does not represent any actual pattern of change in
the nebula's brightness.  No isolated difference feature is visible
which might correspond to rapid, localized changes in brightness at
the location of the gamma-ray flare event.  At the distance of the
Crab (2 kpc), a speed of $c$ corresponds to 2\farcs6 per month.  One
might therefore expect that any feature related to the sudden energy
release of the 2012 July 3 gamma-ray flare would be less than
9\farcs5 in diameter by the second epoch.

We show a detail of the central region in Figure~\ref{fdiffcentre}.
Note the pattern of elliptical ripples visible around the pulsar.  The
pattern is very similar to that seen by \citet{Crab-2001} and
\citet{Crab-2004}, and almost certainly represents real wisp motions
over our 6-day interval. 

The rms differences over the body of the Crab are $\sim$7 times the
rms background value of the difference image.
One might therefore conclude that the differences in the two images
are due to real changes in the nebula's brightness.  As mentioned,
however, no particular feature associated with the gamma-ray release
is visible.  Furthermore, although temporal changes over a large part
of the nebula have been seen in difference images spanning larger time
intervals in the past, \citep{Crab-2004, Crab-2001}, the pattern in
the present difference images does not resemble those earlier
patterns, but rather suggests linear ripples extending over the whole
body of the nebula, which seem physically less plausible.

\subsection{Central Point Source}
\label{scentral}

A faint unresolved source can be seen in the centre of our images
(Figs.\ \ref{fcrabimage}, \ref{fcentre}).  It has a 5.5-GHz flux
density of $1.05 \pm 0.13$~mJy on Aug.\ 20 and $1.17\pm0.16$~mJy on
Aug.\ 26, where the uncertainty includes a contribution due to the
difficulty in reliably separating the point source from the
background.

We determined the position of this source on images made
before any self-calibration (since self-calibration can cause spurious
position shifts), and found that it was within $25 \pm
40$~milli-arcsecond (to the northwest) of the position of the pulsar
given by \citet{LobanovHM2011}.
We conclude that the central point source is coincident with the
pulsar to within 0\farcs14 ($3\sigma$ upper limit).  Its flux density
rose by a not statistically significant amount of $(0.12 \pm 0.21$~mJy
or 0.12 mJy or $(11 \pm 19)$\% over our six-day interval.

\section{A cautionary tale}
\label{scaution}

The differences between our two epochs of Crab radio observations were
about 50$\times$ larger than the off-source image rms.  Are these
differences real?  If so, what is their origin?  The difference image
in Figure~\ref{fdiff2012} suggests striations extending over the whole
body of the nebula, but concentrated at particular position angles and
spatial frequencies, in other words, originating in small regions in
the Fourier transform or \uv-plane.  Such a pattern is suggestive of
errors due to un-sampled regions of the \uv~plane rather than real
changes in the nebula's brightness.

Our difference image consists of the difference of two images, each of
which was made by Fourier-transforming and then deconvolving an
incompletely sampled set of visibility measurements. The deconvolution
process attempts to interpolate into the un-sampled regions of the
\uv~plane in a manner consistent with the known image-plane
constraints.  In the case of our maximum entropy deconvolution, these
constraints are firstly that the true brightness must be positive, and
secondly that it be confined to some region in the original image
(often termed the ``CLEAN'' window even in cases where CLEAN is not
the deconvolution algorithm used.  Note that the images reproduced in
Figs.~\ref{fcrabimage} and \ref{fdiff2012} show only this CLEAN-window
sub-region of the complete images, which latter spanned 18\arcmin\ in
each coordinate).  In \uv~plane, then, any region which was not
sampled is more weakly constrained than those regions which
were sampled.  In the Fourier transform of the {\em difference} image,
therefore, it is those regions which were not sampled in {\em both} of
our two observing epochs which are more weakly constrained.

\begin{figure*}
\centering
\includegraphics[width=0.75\textwidth]{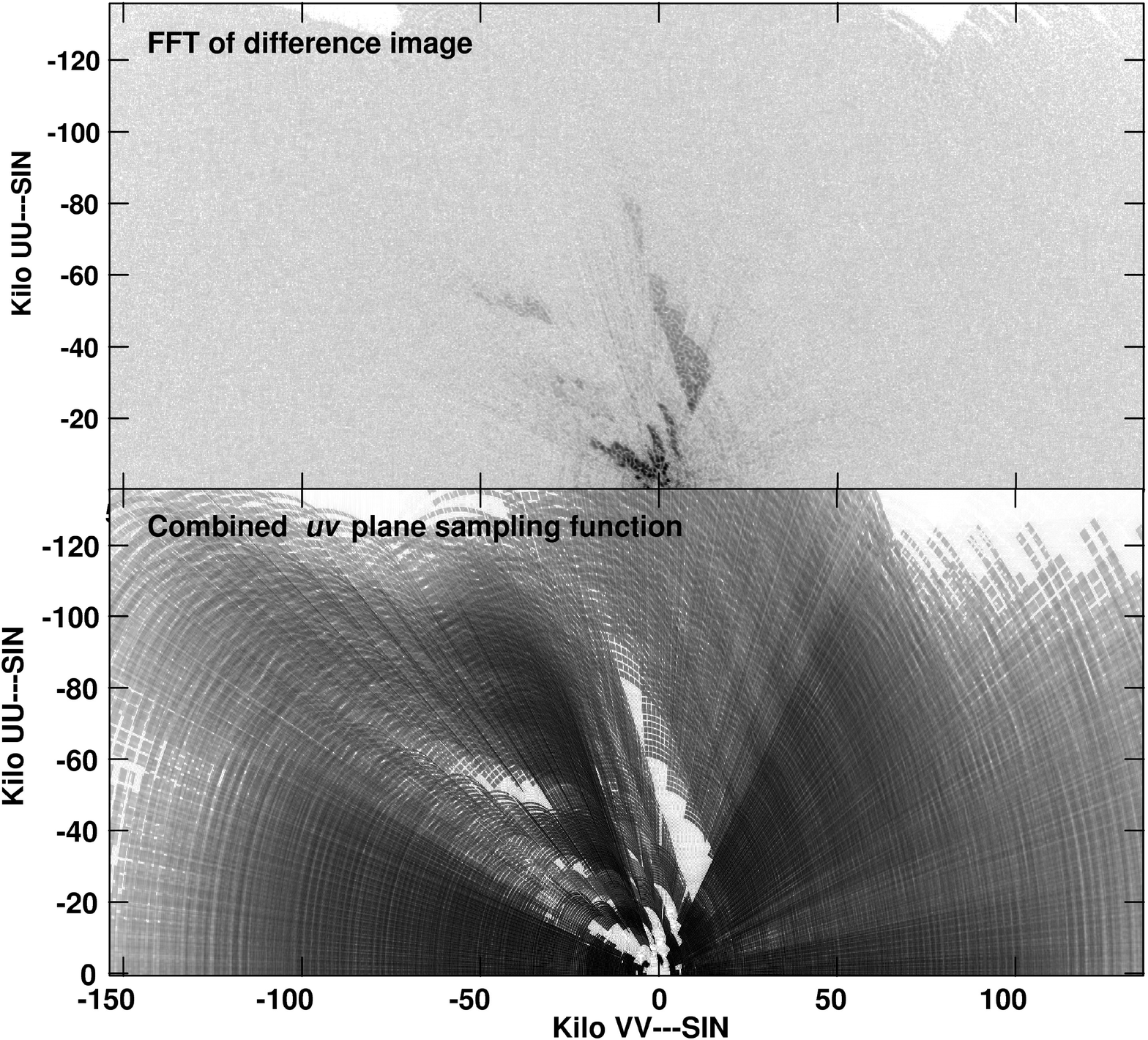}
\caption{{\textbf Top:} The central region of the Fourier transform
  (FFT) of the difference image in Figure~\ref{fdiff2012}, which was
  made by subtracting the 2012 August 20 image from the 2012 August 26
  one.  We show the magnitude of the complex-valued Fourier transform
  of the image, with 0 being white and positive values being dark.
  This shows the regions of the \uv~plane contributing the most power
  to the difference image.  As the Fourier transform of the
  real-valued image is symmetrical, we show only half the \uv~plane.
  {\textbf Bottom:} The same central region of the FFT of the {\em
    product} of the dirty beams corresponding to the two images above.
  The Fourier transform of the dirty beam is essentially the sampling
  function in the \uv~plane, and the product therefore represents the
  combined \uv~plane sampling function.  The blank areas in the
  product of the two sampling functions, then, show the regions of the
  \uv~plane where we did {\em not} have measurements at both epochs,
  and therefore the spatial frequencies where our difference image is
  poorly constrained by the observations.  Note that the {\em
    un-sampled} regions correspond very well with those contributing
  most of the power to the difference image in the top panel.}
\label{fdifffft}
\end{figure*}

We formed the fast Fourier transform (FFT) of the difference image.
It is shown in the top panel of Figure~\ref{fdifffft}.  For each
individual epoch, the beam is the Fourier transform of the sampling
function. The {\em product} of the Fourier transforms of the beams
therefore show the regions which were not sampled by both epochs,
since in such un-sampled regions, the product is near zero.  We show
this product of the \uv~plane sampling functions, in other words the
combined sampling function, in the lower panel of Fig.~\ref{fdifffft}.

It is evident by comparing the two panels of Fig.~\ref{fdifffft} that
the power in the difference image lies predominately at those spatial
frequencies where the combined sampling function was zero.  We can
therefore conclude that the majority of the structure seen in the
difference image, despite being many times larger than the background
rms brightness, is not real, but due only to errors in the
deconvolution, which is necessitated by the incomplete sampling in the
\uv~plane, or as they are commonly termed ``deconvolution errors''.

The rms brightness of the difference image over the body of the Crab
is 306~\uJb\ (at a resolution of 1.10\arcsec $\times$ 1.00\arcsec).
Expressed in terms of the peak brightness, the difference rms is
1.2\%, while the difference extrema are $\sim$7\%.  We conclude that
the deconvolution errors, even in this case of the excellent
\uv-coverage afforded by the VLA wide-band system and several hours
of observations, are at an rms level of 1.2\% of the image brightness,
with extrema at 7\%.  The off-source rms in these images was
$\sim$30~\uJb\
so the deconvolution errors are larger than the off-source noise level
by an order of magnitude!  We made a similar difference image using
CLEAN rather than maximum entropy deconvolution, and found that the
deconvolution errors were larger by $\sim$30\% than they were for the
maximum entropy deconvolution.  CLEAN is known to perform more poorly
on extended sources \citep{SaultO1996, NarayanN1986}, so the fact that
CLEAN deconvolution errors are higher than those obtained from maximum
entropy is expected for an extended object like the Crab.

The effective uncertainty in the brightness of the image is thus much
larger than the noise, and the image is strongly dynamic range
limited.  We note that the fractional errors are probably smaller in
the case of spatially smaller sources, and for single unresolved
sources the image error will likely approach the thermal noise.

\section{Differences over longer timescales}
\label{slongdiff}

On the one-week timescale between our two 2012 images, we found above
that differences were small and only in the centre of the nebula were
convincing changes corresponding to rapid motions seen.  However,
slower motions are known to occur both in the centre and over
the body of the nebula \citep{Crab-2004, Crab-2001, Crabwisp-1992}.
We compare our new images to those made using VLA observations in
2001, also at 5~GHz and using the B array configuration.  The 2001
observations are described in \citet{Crab-2004}.

One goal of this comparison is to measure the expansion rate of the
nebula. We defer discussion of the measured expansion rate to a future
paper.  In order to avoid biases introduced by scaling the default in
the deconvolution with the expansion rate measured by
\citet{Crab-expand}, we image using only the B array configuration
data without using any default or support.  As a consequence, our
images are hi-pass filtered, and may include genuinely negative
values.  Since maximum entropy deconvolution enforces positivity and
therefore cannot deal with true negative values in the images, we turn
to CLEAN deconvolution.  To avoid spurious differences due to
differing \uv~coverage either at the short or long end of the spacings
covered by the B configuration, we further hi-pass filtered both
images with a Gaussian of FWHM 20\arcsec, and also convolved both to a
common restoring beam of $2.00\arcsec \times 1.80\arcsec$ at p.a.\
80\arcdeg.

The flux scaling and registration in RA and Dec.\ might be slightly
discrepant between our two images, which were made using
self-calibration.  Such discrepancies would have a significant effect
on difference images.  We therefore used the Miriad program IMDIFF,
which calculates how to make one image most closely resemble another
by calculating unbiased estimators for the scaling in size, $e$, the
scaling and the offset in flux density, $A$ and $b$ respectively, and
the offsets in RA and Dec, $x$ and $y$ respectively, needed to make
the second image most closely resemble the first.

\begin{figure*}
\centering
\includegraphics[width=0.9\textwidth]{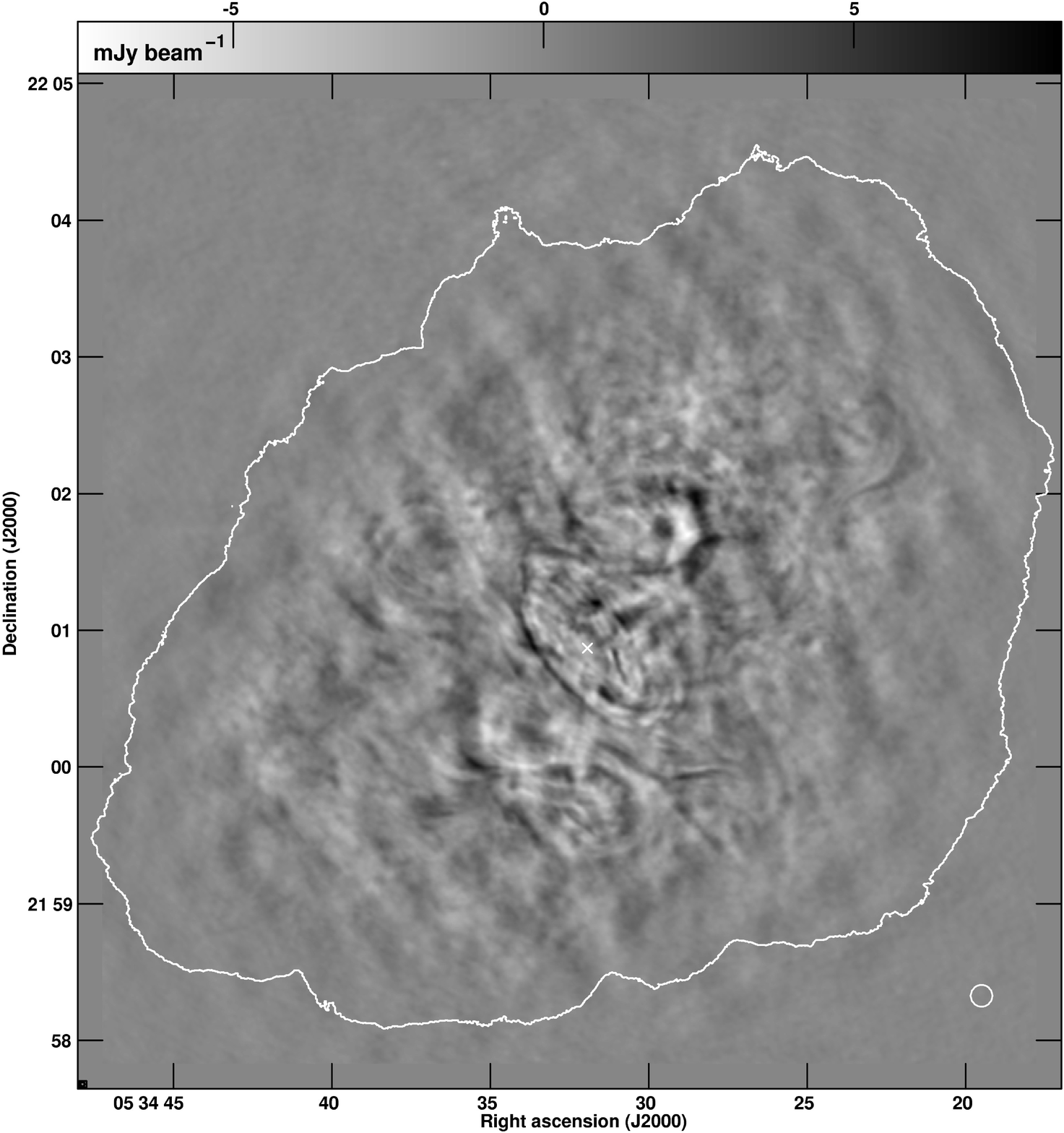}
\caption{A difference image of the Crab, showing the changes between
  2001 April 16 and 2012 August 26\@ (with the former subtracted from
  the latter).  The difference image has been high-pass filtered at
  20\arcsec\ to isolate structures well sampled by the B-array which
  was used for both sets of observations.  Both were convolved to the
  same FWHM resolution of $2.00\arcsec \times 1.80\arcsec$ at
  p.a.\ 80\arcdeg.  (Note that the
  greyscale range is much larger than in Fig.\ \ref{fdiff2012}).  As
  in Fig.\ \ref{fdiff2012}, the white cross shows the position of the
  pulsar, the 1\% contour of the 2012 image is indicated, and the
  circle at lower right shows a diameter of 9\farcs5, which is the
  diameter that a feature expanding isotropically with speed $c$ since
  the gamma-ray flare onset on 2012 July 3 would have in 2012 August 26.
  Before subtraction, the 2001 image was expanded by a factor of 1.016
  to account for the expansion of the nebula (see text for details).}
\label{fdiff2012-2001}
\end{figure*}

Finally we show in Figure~\ref{fdiff2012-2001} the {\em residual}
image after finding the best fit values of $e, A, b, x$ and $y$ above
We treat $e$ here merely as a nuisance parameter and defer the discuss
the expansion rate to a future paper. We discuss here the slow but
significant changes in the radio brightness over the whole body of the
nebula not attributable to the overall expansion.  This difference
image shows the changes that have occurred in the Nebula over the
11.4~yr period between 2012 and 2001 (at spatial scales between
$\sim$2\arcsec\ and 20\arcsec).

The changes in the brightness over the 11.4~yr interval between 2001
and 2012 involve a complex series of ripples.  The extrema in the
difference image are $-7.50$ and +8.34 m\Jb.  The rms over the nebula
is 1.20~m\Jb, while the rms in a region 1\arcmin\ in radius around the
pulsar is 1.97~m\Jb\ (FWHM beam area = 4.08~\invassq).  By comparison
the rms variation between the two 2012 epochs (Fig.\ \ref{fdiff2012})
were four times smaller with rms 0.30~m\Jb\ over the nebula (albeit
with a smaller FWHM beam area of only 1.25 \invassq). Also, comparing
the pattern of differences the pattern visible of the 11.4-year
interval are much less regular, and a FFT of the 11.4-year difference
image shows no patterns related to the \uv~coverage.  We conclude
therefore that most of the changes visible in
Fig.\ \ref{fdiff2012-2001} are real, although this difference image
will also contain artifacts due to deconvolution errors.

\section{Discussion}
\label{sdiscuss}

In response to a gamma-ray flare in the Crab nebula in 2012 July, we
observed the Crab Nebula using the VLA with $\sim$1\arcsec\
resolution (B-array configuration) in the hopes of seeing a radio
counterpart to the gamma-ray flare which would allow us to accurately
localize the flare.  We observed the nebula twice, 49 and 55 days
after the onset of the gamma-ray flare.  We found that there were no
rapid changes in the radio emission (at 5.5 GHz) from the nebula which
could be attributed to the gamma-ray flare over our 6-day interval.
The extrema in the difference image were $\sim$1.5~m\Jb, or 5\% of the
peak brightness in the nebula, however, we show that most of these
difference are in fact artifacts due to deconvolution errors.

The rapid variability of the gamma-ray flares show that the
acceleration region must be quite small, with a radius of $\lesssim
10^{-4}$~pc \citep{BuehlerB2014}, corresponding to $\lesssim
$0\farcs01 at 2 kpc.  The acceleration zone for the flares would
therefore be completely unresolved in our image.  Any features
associated with the flare can expand at most at $c$, it must therefore
have a radius of $\lesssim 55$~light-days, corresponding to
4.8\arcsec\ by the time of our August 26 observations.  No such
feature is seen in our difference images.  Therefore we can say that
any radio feature associated with the gamma-ray flare is changing in
surface brightness
by $\lesssim 0.2$~m\Jyas\ per day at 5 GHz.

We compare our radio images also to an older one from 2001 April, when
there was likely no gamma-ray flare.  Since the Fermi satellite was
launched only on 2008, no gamma-ray observations were available, but
since flares seem to have a lifetime on the order of a week and occur
on the order of twice per year, the chances that the Crab was flaring
during the 2001 April observations are a few percent.  Again, no
particular compact feature, where the radio emission in a region
$<9\farcs5$ in diameter has brightened, stands out, although we note
that there are changes throughout the nebula.  The extremum in the
2012 -- 2001 difference image is at 2~mJy~arcsec$^{-2}$,
so we can say that any feature associated with the flare had a
5-GHz brightness of less than 2~mJy~arcsec$^{-2}$, or a total of $<145
$~mJy (assuming a maximal radius of 55 light-days), corresponding to a
spectral luminosity of $7 \times 10^{20}$~erg~s$^{-1}$~Hz$^{-1}$.

We now discuss the radio emission from specific regions in the nebula,
and whether or not they might represent a radio counterpart of
the gamma-ray flares.

\subsection{Radio Emission from the Pulsar}
\label{spulsar}

We found an unresolved, likely variable, source with flux density
$\sim$1~mJy in our images which was coincident with the pulsar.  Is
this just the pulsed radio emission from the pulsar, averaged over the
pulse (period = 33~ms)?  

Unfortunately the pulse-averaged flux density of the pulsar is not
well known.  \citet{MoffettH1996} found that the pulsar's flux density
at 5~GHz is quite variable, and give a mean value of 0.57~mJy for
4.9~GHz.  \citet{Cordes+2004a} also find that above 3~GHz it is quite
variable on timescales as short as five minutes.  We examined the four
4.8-GHz VLA images of \citet{Crab-2004}, taken over a period of two
months in 2001, and found that the pulsar cannot be reliably
identified in them, suggesting that its flux density is $<2$~mJy.
However, the brightness at the location of the pulsar varies with
0.4~mJy rms over the four images, suggesting that the pulsar's flux
density is variable by that amount over times scales of a few
weeks. The pulsar was also detected in two 8.4~GHz VLA images with
sub-arcsecond resolution taken on 1991 July 14 and Oct.\ 30
(D. A. Frail and M. F. Bietenholz, unpublished), and had flux
densities of $\sim$0.7 and $\sim$0.1~mJy, respectively.
\citet{LobanovHM2011} give an upper limit of 0.4~mJy to the pulsar's
emission at 4.9~GHz in 2010 Nov.\ 23.  Although it is unknown whether
or not a gamma-ray flare had occurred prior to these earlier
observations, it is unlikely since flares seem to occur once or twice
per year.  Taken together these results suggest that the
pulse-averaged emission from the pulsar is quite variable, with a mean
value at 5 GHz in the range of 0.5 to 1 mJy, and an rms variability of
at least 0.4 mJy, with timescales possibly as short as a few minutes.

The flux densities for the central point source we observed in 2012
are therefore within the range of the normal radio emission from the
pulsar.  Since the point source is coincident with the pulsar to
better than 0\farcs14 ($3\sigma$), corresponding to $4.2 \times
10^{15}$~cm or $\sim 10^{7.4}$ light-cylinder radii, 
the most parsimonious explanation of the unresolved point source in
our images is that it is the pulse-averaged emission from the pulsar
itself, and is not directly connected to the gamma-ray flare.

We note that \citet{Weisskopf+2013} report a detection of emission
coincident with the pulsar at up to 5.0~mJy at
7.8~GHz\footnote{Averaged between the values at 7.762 and 7.872 GHz
reported in \citet{Weisskopf+2013}.} and 1.78~mJy at 4.9~GHz.  The
unusually high value of 5.0~mJy was observed only $\sim$17 days after
the onset of the gamma-ray flare.  It is large, but not impossibly so
for being the normal emission from the pulsar.

\subsection{The Inner Knot} 
\label{sinnerknot}

There is a compact, variable optical and infra-red
feature close to the pulsar called the inner knot, located $\sim
0\farcs65$ from the pulsar at p.a.\ 120\arcdeg\ \citep{Hester+1995,
  SandbergS2009}.  It is variable in brightness\citep{SandbergS2009},
and using Hubble Space Telescope images, \citet{Sollerman2003} found
the position of this feature to also be variable at the 0\farcs1
level.  None of the reported optical or infra-red positions of the
inner knot, however, are closer than 0\farcs6 to the pulsar.
The inner knot is somewhat extended, with \citet{Hester+1995}
reporting an angular size of 0\farcs16 in the direction to the pulsar
and about three times larger in the perpendicular direction.

\citet{KomissarovL2011} suggested that the inner knot might be the
site of the gamma-ray flaring activity on the basis of its proximity
to the pulsar and the short synchrotron lifetimes of the gamma-ray
emitting electrons.

The central radio point source discussed in \S~\ref{scentral} above is
therefore not associated with the inner knot since the $3\sigma$ upper
limit on the displacement of the central radio point source from the
pulsar is only 0\farcs14.

We estimated an upper limit on flux density of the inner knot by
fitting and subtracting the pulsar from the image shown in
Figure~\ref{fcentre}.  From this residual image we estimate that the
largest feature which might escape detection at the location of the
inner knot is 0.3~m\Jb.  We therefore find an upper limit
to the 5.5-GHz radio emission from the inner knot of 0.3~m\Jb.
If the inner knot is, as suggested by \citep{KomissarovL2011},
the site of the gamma-ray flaring activity, there is very
little corresponding radio activity.

In the infrared, \citet{Weisskopf+2013, SandbergS2009, Melatos+2005}
and \citet{Sollerman2003} all report flux densities for the inner
knot.  The average of the inner knot flux densities reported in the
infrared K band ($1.4\times10^{14}$~Hz) from those authors is
0.36~mJy. Although \citet{Weisskopf+2013} report that the K$^\prime$
infrared flux rose by 35\% following a strong flare, this rise is
compatible with the normal variation of the inner knot, and therefore
likely not associated with the flare.  In the radio we did not detect
the inner knot, and therefore also find no evidence that the inner
knot shows any response to gamma-ray flares.

Our upper limit on the 5.5-GHz flux density of the inner knot along
with the infrared values implies a radio-infrared spectral index,
$\alpha_{\rm Rad}^{\rm IR}> +0.02$.  By contrast, in the infrared,
\citet{Melatos+2005} and \citet{Sollerman2003} report that the inner
knot's spectral index is $\sim -0.8$ while \citet{SandbergS2009}
suggest a value of $-1$.  The spectral energy distribution of the
inner knot therefore must have a turnover somewhere below
$10^{14}$~Hz. The knot's $\alpha_{\rm Rad}^{\rm IR}$ must be rather
flatter than the integrated value for nebula, which is $\sim -0.4$
\citep[e.g.][]{AharonianA1998, Arendt+2011}.

There is another constraint that can be drawn from our present
observations and those of \citet{LobanovHM2011} and this is from the
limit that can be placed on the flux from a point source at the
position of the inner knot.  The inner knot has been associated with
the region where the pulsar outflow is deflected towards us by a
standing termination shock \citep[e.g.][]{KomissarovL2011}.  The shock
should be located where the momentum flux in the wind, thought to
scale as the cosine of the latitude to the fourth power
\citep{TchekhovskoySL2013}, matches the ambient nebula pressure. If
the momentum flux is scaled to the inner edge of the equatorial torus
seen in X-rays then the shock radius along the line of sight can be
estimated as $\sim$9\arcsec\ in angular measure.
In this case the deflection angle is $\sim$0.07~rad and the
dissipation would have to be substantial. The 5.5~GHz flux density
limit on the inner knot is $\sim 300 \, \mu$Jy, which implies $\nu
L_\nu$ per steradian at 5.5~GHz of $6 \times 10^{26}$
erg~s$^{-1}$~sr$^{-1}$ or $1.6 \times 10^{-11}$ of the spindown
luminosity of the pulsar per steradian.

This surprisingly low radio radiative efficiency of $\lesssim 1.6
\times 10^{-11}$ is a strong constraint on models of the termination
shock which should produce some radio synchrotron emission from
$\sim$~GeV electrons as well as the low harmonics radiated by more
energetic particles. In addition shocks can create coherent radio
emission associated with the surface current at the shock front and
this, too, will be limited.

\subsection{Other Radio Counterparts to Gamma-Ray Flares}
\label{scounterpart}

Our observations did not detect any radio emission which could be
identified as a radio counterpart to the gamma-ray flare.  Earlier
attempts to find such radio emission by \citet{LobanovHM2011} and
\citet{Weisskopf+2013} had also proved inconclusive.  Unlike some of
the earlier observations, we are sensitive to emission anywhere in the
body of the nebula, not just near the pulsar.  Our observations are
also more sensitive to somewhat more extended emission at late times,
as might occur if the source were expanding relativistically.

The 1.6-GHz VLBI observations carried out 44~d after a gamma-ray flare
in 2011 September did result in a possible detection of radio features
associated with that gamma-ray flare: \citet{LobanovHM2011}
tentatively identified two compact knots of radio emission, which they
termed C1 and C2, as possible sites of the flaring activity.
We indicate the two positions of C1 and C2 on 
our Figs.~\ref{fcentre} and \ref{fdiffcentre}. 
No particular feature or change is visible at the position of either
of these knots.  \citet{LobanovHM2011} report 1.6-GHz flux densities
of $0.5 \pm 0.3$ and $0.4 \pm 0.2$ mJy for the two knots respectively.
On our 5.5-GHz image, no corresponding features can be seen down to a
level of $\sim$0.4 mJy.  We therefore cannot confirm the radio
emission reported by \citet{LobanovHM2011}.  \citet{Weisskopf+2013}
also report on radio observations following the strong flare of 2011
April, and report no radio emission features at the locations of C1
and C2 with the best $3\sigma$ upper limits on the order of 0.5 mJy.
They did not detect any significant variability in the X-ray emission
at these locations either.

More generally, we can say that the 5.5-GHz spectral luminosity of any
feature associated with the flare in our observations was $\lesssim 2
\times 10^{-4}$ that of the nebula.  By contrast, even a relatively
weak gamma-ray flare like the one of 2012 July involves an increase of
factors of several of the nebular flux above 100~MeV.  In other words
whatever the process involved in the flares is, they are faint in the
radio \citep[this conclusion was already reached in
  e.g.][]{Weisskopf+2013}.

The total energy of the 2012 July flare was of order $10^{41}$~erg.
Given an average magnetic field in the nebula of 300~$\mu$G, this
energy is the equivalent of the magnetic energy contained in a volume
of radius $\sim 2 \times 10^{16}$~cm (0.006 pc).

The energy released in the gamma-ray flares is substantial, with the
flare fluxes being up to 1\% of the pulsar's spindown energy loss
rate, and the flare timescales implying that this energy release
happens in very small regions.  It seems likely that the dynamical
aftermath of such an energy release would affect the ambient emission,
for example through compression or rarefaction of the magnetic field
and/or the particle density \citep{Weisskopf+2013}.  Such disturbances
would propagate through the synchrotron plasma filling the bulk of the
nebula, likely at the sound speed of $\sim c/\sqrt{3}$.  The temporal
changes that we observed may therefore be the ripple-like disturbances
created in the synchrotron-emitting fluid by the sudden energy
releases which produce the gamma-ray flares, propagating through the
nebula and interacting with the complex structure of the dense thermal
filaments and/or the outer boundary of the nebula.  Alternatively, the
temporal changes may similarly be the propagating, wave-like effects
of instabilities at the termination shock, which instabilities
manifest themselves in the rapid variability of the wisps.
The rate of rotational energy loss by the pulsar is prodigious at
$\sim 5 \times 10^{38}$~erg~s$^{-1}$ \citep[e.g.][]{Hester2008}, and
most of this energy is thought to be injected into the nebula in the
form of magnetic energy \citep[e.g.][]{PorthKK2014, BuehlerB2014,
  ReesG1974}.  Since the majority of this magnetic energy must be
dissipated on timescales less than the $\sim$1000-yr age of the
nebula, variation in emissivity on scales from hours to decades is
probably not surprising.

The radio-emitting electrons should evolve adiabatically as their
synchrotron cooling times are very long compared with the timescales
for variation. In this case the synchrotron emissivity at frequency
$\nu$ varies $\propto B^{5/2-2\alpha}\nu^{\alpha} \propto
B^{3.04}\nu^{-0.27}$, where $B$ is the magnetic flux density and we set
the radio spectral index $\alpha=-0.27$ \citep[e.g.][and references
  therein]{Crab-1997}. Suppose first that the $\gamma$-ray emission
from a large flare is due to an electromagnetic interchange that
releases energy $E\sim10^{42}E_{42}$~erg. The ambient energy density
in the central
regions of the nebula is $\sim10^{-7}$~erg cm$^{-3}$ and so we might
expect to obtain $O(1)$ changes in the magnetic field and the
synchrotron emissivity over a region of size
$\ell \sim 10^{16}E_{42}^{1/3}$~cm $\sim 0.5 \, E_{42}^{1/3}$~arcsec
$\sim 5 \, E_{42}^{1/3}$~light days. If we take the effective
line-of-sight depth of the nebula to be equivalent\footnote{The FWHM
  of the nebula in the sky plane is $\sim$200\arcsec\ at 5.5~GHz, and
  we take this value as an estimate of its line-of-sight depth.}
to 200\arcsec\ and the central diffuse surface brightness to be
$\sim$15~m\Jyas\ (see Fig.\ \ref{fcrabimage}, \ref{fcentre}),
then the expected surface brightness change on scale $\ell$ is $\sim
150 \, (\ell/ 1{\rm arcsec})\,\mu$\Jyas, or $\sim 75 \, E_{42} \,
\mu$\Jyas.  Since our beam area is $\sim$1~\invassq\ and is comparable
to or larger than the expected values of $\pi \ell^2$, we would in
fact only expect changes smaller than $\sim 75 \, E_{42} \,
\mu$\Jb\ to appear in our images.  Furthermore, for the particular
flare of 2012 July, $E_{42}$ was only of order 0.1, so the expected
changes in the radio brightness due to the flare would be too small to
be detected, especially against the background of radio variability in
the nebula that is not directly connected with the flare.

These estimates of the expected change in the radio emission due to a
flare are quite model-dependent and do not take account of
beaming. Although the limits on both the surface brightness and the
flux discussed above are therefore not, in practice, very
constraining, they do support the view that the gamma ray flares have
a fairly high radiative efficiency.

\subsection{Filamentary structure and temporal variations in the Nebula}
\label{sfilaments}

Our radio image in Figure~\ref{fcrabimage} shows prominent filamentary
structure visible throughout the nebula.  Such structure has been
seen in earlier radio images \citep[e.g.][]{Crab-2004, Crab-2001}.  At
our resolution of $\sim 1\farcs0$, most of the filamentary structure
seems resolved.

The filamentary structure visible in our 5.5~GHz is generally well
correlated with that at other radio wavelengths from 74~MHz
\citep{Crab-1997} to 350~GHz \citep{Arendt+2011, GreenTP2004},
although the images available at these other frequencies are of lower
resolution.
At long infrared wavelengths the filamentary structure still largely
corresponds to that visible in radio, but at shorter infrared
wavelengths \citep{Temim+2006} and in optical \citep{Loll+2013} much
of the filamentary structure has disappeared.  Since the optical
continuum emission is synchrotron emission like the radio, it might be
expected to have a similar morphology, but instead shows much less
filamentary structure.  A possible reason for this difference may just
be synchrotron burnoff.  Much of the filamentary structure resides in
the outer nebula, and if all the relativistic particles originate in
the termination shock near the pulsar, then the synchrotron lifetimes
of the higher-energy electrons are too short to reach, and therefore
illuminate in the optical, the outer filaments \citep{TangC2012}.  It
has also been proposed, however, that the radio-emitting particles are
accelerated in the vicinity of the filaments, which would also be
consistent with the observed decrease of the filamentary structure
towards the optical.

Furthermore, although we did not find any particular radio variability
which might be associated with the gamma-ray flare, we did find that
over a period of about a decade there were striking changes in the
radio emission occurring throughout the nebula, which were most
pronounced within $\sim$1~pc of the pulsar.  Such changes had
previously been reported \citep{Crab-2004, Crab-2001, Crabwisp-1992},
but are now more clearly visible.  The temporal variations were up to
10\% of the peak brightness of the nebula (at $\sim$1\farcs9
resolution).  The changes have a complex morphology of arcs and knots:
there is little structure at the $\sim$2\arcsec\ level, but
considerable structure on spatial scales of a few arcsec
($\sim$0.04~pc) up to perhaps 1\arcmin\ ($\sim$0.6~pc).  Earlier,
\citet{Crab-2004} had examined a differences in radio image made
$\sim$3~yr apart, and found similar changes, but which were mostly
confined to a region within $\sim$1\arcmin\ from the pulsar.  Over our
longer timescale of about a decade there are significant changes at
least 2\arcmin\ from the pulsar.  This suggests that the variability
timescale becomes longer at larger distances from the pulsar.

We note that exactly such a pattern of radio brightness changes is
seen in the dynamical relativistic magneto-hydrodynamic (MHD) model of
\citet{Olmi+2014}.  Their model was axisymmetric and had parameters
chosen to best reproduce the Crab's X-ray morphology, but unlike most
previous MHD modelling, Olmi et al.\ examined the morphology of the
resulting radio rather than the high-energy emission.  They find that
the radio morphology is essentially the same regardless of whether the
radio-emitting electrons are continuously accelerated at the
termination shock or if they are uniformly distributed in the nebula.
The radio morphology (including the radio wisps) therefore reflects
only the underlying flow structure, but not the site where the
radio-emitting electrons are accelerated. In their model the injected
magnetic field from the pulsar changes polarity around the rotational
equator, and eddies then cause the current sheet to twist and tangle
downstream of the termination shock, and it is these instabilities
which give rise to the observed variability in the radio brightness,
and produces synthesized radio difference images very much like the
observed one in Figure~\ref{fdiff2012-2001}.  In agreement with our
observations the modelled radio brightness changes in the inner nebula
are more rapid and of larger amplitude than those in the outer nebula.
Our observations therefore lend considerable credence to the MHD
modelling of \cite{Olmi+2014}.

\section{Conclusions and summary}
\label{sconclude}

We briefly summarize our results and conclusions:
\begin{trivlist}

\item{(1)} We present new deep 5.5-GHz radio images of the Crab Nebula,
  about two months after a gamma-ray flare event in 2012 July.  

\item{(2)} We find no significant change in the Crab's radio emission
  localized to a region $<2$ light-months in radius.  Any radio
  counterpart to the gamma-ray flare has a total flux density of
  $<145$~mJy, corresponding to a spectral luminosity of $<7 \times
  10^{20}$~erg~s$^{-1}$~Hz$^{-1}$, or $< 2 \times 10^{-4}$ that of
  the nebula.  The surface brightness of any flare counterpart is $<
  2$~mJy~arcsec$^{-2}$. 

\item{(3)} The low limits for a radio counterpart to the flare imply
  that few low-energy electrons ($\lesssim 10$~GeV) are accelerated in
  the flaring event.  Nonetheless, the energy release of the flare is
  expected to produce changes in the magnetic field, which should
  produce corresponding changes in the radio brightness.  We show
  however, that such changes are likely to be smaller than our
  observed radio brightness limits on energetic grounds.

\item{(4)} We detect radio emission at the mJy level from within
  0\farcs14 ($4.2 \times 10^{15}$~cm) of the pulsar.  This emission is
  very likely just the normal pulse-averaged emission from the pulsar,
  but we cannot rule out mJy-level radio emission associated with the
  gamma-ray flare from very near the pulsar at this level.

\item{(5)} We find no discernible radio emission from the ``inner
  knot'', seen at $\sim 0\farcs65$ from the pulsar in the optical and
  infra-red.  We set an upper limit of 0.3~m\Jb\ on the 5.5-GHz radio
  brightness of the inner knot.  This limit represents a very low
  radiative efficiency of $1.6 \times 10^{-11}$ of the spindown
  luminosity of the pulsar per steradian.

\item{(6)} We find that deconvolution errors are several times larger
  than the thermal noise, even in these images made using the wide
  bandwidth and consequently excellent \uv~coverage of the VLA.  The
  deconvolution errors represent a fraction of 1.2\% of the peak
  brightness of the nebula, although this fraction is likely strongly
  dependent on the $\uv$~coverage and source geometry. 

\item{(7)} By comparing our images to ones from 2001, we find
  widespread changes in the brightness over decade timescales, as
  large as 2 mJy~arcsec$^{-2}$, or up to $\sim$10\% of the peak
  brightness of the nebula.  These changes are both larger in
  amplitude and morphologically distinct from the deconvolution
  errors.  Averaged over the nebula the changes in surface brightness
  over decade timescales have an rms of 1.4\% of the peak brightness.
  (These changes are in addition to the secular decay in brightness of
  $\sim$1.3\% per decade).  The morphology of the changes is complex,
  suggesting both filamentary and knotty structures.  The variability
  is stronger in the centre of the nebula, and the timescales are
  likely shorter near the centre than at the periphery.
  These variations correspond well to those seen in MHD
  simulations of the Crab by \citep{Olmi+2014}.

\end{trivlist}

\section*{Acknowledgements}

We are indebted to the late Dr. Michael Gaylard for his encouragement
and support of astronomical research at Hartebeesthoek Radio Astronomy
Observatory.  We also thank David T. Thompson for comments on the
manuscript, and Jon Arons, Martin Weisskopf and Claire Max for support
and encouragement.  We thank the anonymous referee for useful comments
on the manuscript.  Research at Hartebeesthoek was partly supported by
National Research Foundation (NRF) of South Africa.  Research at York
University was partly supported by NSERC\@. R. Blandford acknowledges
support from NSF grant AST 1212195\@.  The National Radio Astronomy
Observatory is a facility of the National Science Foundation operated
under cooperative agreement by Associated Universities, Inc.  We have
made use of NASA's Astrophysics Data System Bibliographic Services.

\bibliographystyle{mn2e}  

\bibliography{mybib1}
\end{document}